\documentclass{sig-alternate}

\usepackage{color}
\usepackage{graphicx}
\usepackage{subfig}

\newcommand{\superscript}[1]{\ensuremath{^{\textrm{#1}}}}
\def\sharedaffiliation{\end{tabular}\newline\begin{tabular}{c}}
\def\wu{\superscript{\S}}
\def\wg{\superscript{\dag}}

\begin{document}
\setlength{\pdfpagewidth}{8.5in}
\setlength{\pdfpageheight}{11in}

\conferenceinfo{WSDM'13,} {February 4--8, 2013, Rome, Italy.}
\CopyrightYear{2013}
\crdata{978-1-4503-1869-3/13/02}
\clubpenalty=10000
\widowpenalty = 10000

\title{Distinguishing Topical and Social Groups Based on Common Identity and Bond Theory}

\numberofauthors{4}
\author{
\alignauthor \hspace{-0.5cm} Przemyslaw A. Grabowicz\wu\titlenote{P.A.G. has performed the research while visiting Yahoo! Research Barcelona.}\\
\alignauthor \hspace{-1.5cm} Luca Maria Aiello\wg\\
\alignauthor \hspace{-3.0cm} V\'{\i}ctor M. Egu\'{\i}luz\wu\\
\alignauthor \hspace{-4.5cm} Alejandro Jaimes\wg\\
\sharedaffiliation
{}
  \begin{tabular}{ccc}
  {\wu}\{pms,victor\}@ifisc.uib-csic.es & & {\wg}\{alucca, ajaimes\}@yahoo-inc.com \\
	\affaddr{IFISC (CSIC-UIB)}		&	&	\affaddr{Yahoo! Research}\\
	\affaddr{Palma de Mallorca, Spain}		&	&	\affaddr{Barcelona, Spain}\\
  \end{tabular}
}

\maketitle

\begin{abstract}

Social groups play a crucial role in social media platforms because they form the basis for user participation and engagement. Groups are created explicitly by members of the community, but also form organically as members interact. Due to their importance, they have been studied widely (e.g., community detection, evolution, activity, etc.). One of the key questions for understanding how such groups evolve is whether there are different types of groups and how they differ. In Sociology, theories have been proposed to help explain how such groups form. In particular, the common identity and common bond theory states that people join groups based on identity (i.e., interest in the topics discussed) or bond attachment (i.e., social relationships). The theory has been applied qualitatively to small groups to classify them as either topical or social. We use the identity and bond theory to define a set of features to classify groups into those two categories. Using a dataset from Flickr, we extract user-defined groups and automatically-detected groups, obtained from a community detection algorithm. We discuss the process of manual labeling of groups into social or topical and present results of predicting the group label based on the defined features. We directly validate the predictions of the theory showing that the metrics are able to forecast the group type with high accuracy. In addition, we present a comparison between declared and detected groups along topicality and sociality dimensions.

\end{abstract}

;


\section{Introduction}\label{sec:intro}

A longstanding theory about the creation of social communities affirms that people join groups driven by either strong personal ties with other members or by the interest in the group as a whole. As a result, depending on the prevalent motivation of members, spontaneously generated groups can be categorized as either \textit{social} or \textit{topical}. This theory is known as ``common identity and common bond''~\cite{Prentice94asymmetries} and assumes that the two types of groups have distinct and well recognizable traits that characterize them.

In recent years, the theory has been widely commented and elaborated by social scientists from a theoretical perspective and through small-scale experiments~\cite{Sassenberg02common,Utz02distributive,Ren07Applying}, but a validation over large-scale datasets together with the development of rigorous, automated methodologies to distinguish the group types is missing. Indeed, the availability of big data from social media platforms provides the opportunity to study the dynamics of the online communities from a data-mining perspective~\cite{Mislove2007Measurement,Negoescu08Analyzing,Kairam12Life}. None of those experiments, however, have been directly aimed at verifying the common identity and common bond theory.

The design of a technique to detect the group type based on the common identity and common bond principles would first contribute to a strong validation of the theory itself and, on the other hand, would provide a general framework for automatic classification of user groups in online social media. Such a classification would have direct impact on the ability of describing the structure of online social networks along the axes of sociality and topicality and, consequently, on the possibility of better user characterization.

We contribute to fill this gap by proposing a set of general metrics based on the theory. We show that the metrics' values computed on a large corpus of groups extracted from Flickr confirm the cardinal points of the theory and are indeed good predictors of the group type. In addition, we repeat the same analysis on groups identified by a graph-based community detection algorithm. This allows us to compare the user-generated communities to the automatically detected ones not only from a structural perspective but also along the dimensions of sociality and topicality. Since community detection techniques have been largely employed in recent years to describe the structure of complex social systems~\cite{Fortunato2010Community}, the need for a clearer assessment of the meaning of the detected clusters has been often expressed from different angles~\cite{lancichinetti08benchmark,jaewon12defining}, but never completely satisfied. With our study we also contribute to shed light on this matter.

To the best of our knowledge, this is the first attempt of formalization of the common identity and common bond theory, and of its validation over a large and diverse set of user communities. The obtained results can open a new perspective on the semantic interpretation of implicit and explicit user groups in social media.

Our main contributions can be summarized as follows:
\begin{itemize}
\item Translation of the common identity and common bond theory into general metrics applicable to social graphs. An insightful characterization of a large group dataset from Flickr is performed using the proposed metrics.
\item Comparison between user-defined groups and groups discovered by a community detection algorithm, both in terms of their overlap and their properties of sociality and topicality.
\item Design of a method to predict whether a group is social or topical, based on the defined metrics. Prediction on the user-generated groups from the Flickr data yields surprisingly good results.
\end{itemize}

The rest of the paper is structured as follows. In Section~\ref{sec:related} we overview related work. In Section~\ref{sec:socialtheory} we describe the main principles of the common identity and common bond theory and propose their practical translation into general metrics, and in Section~\ref{sec:dataset} we introduce the Flickr dataset we use to test these metrics. In Section~\ref{sec:labeling} we describe the process of gathering a ground truth about group topicality or sociality through an editorial process. In Section~\ref{sec:analysis} we provide a comparison between user-created Flickr groups and groups identified by a community detection algorithm in terms of their overlap, then we study the properties of the defined metrics computed over declared and detected groups and in particular over the subsets of groups labeled as social or topical in the ground truth that we produced. Finally, in Section~\ref{sec:detection} we show that we can accurately predict the group type by combining the defined metrics.

\section{Related work}\label{sec:related}

The spontaneous aggregation of people in communities has been widely studied in social science as one of the fundamental processes driving the global dynamics of social ecosystems~\cite{mcmillan86sense} and as a factor driving the formation of the social identity of individuals~\cite{Tajfel82Social}.

Since the emergence of online social media, the global structure, evolution and dynamics of social groups have been investigated over large-scale and heterogeneous datasets~\cite{Grabowicz2012Dynamics}. Evolution of groups has been characterized as a very broad phenomenon~\cite{Mislove2007Measurement,Cox11Developing} that is dependent on the nature of the group~\cite{Cummings2002quality}, its intrinsic fitness~\cite{Grabowicz2012Heterogeneity} and on the density of social ties connecting its members~\cite{Backstrom2006Group}. Dependency of activity and connectivity on group size has been studied in several platforms~\cite{Grabowicz2012Social,Kairam12Life,Goncalves11Modeling}, showing relations to Dunbar's theory on the upper bound of around 150 stable social relationships for an average human~\cite{dunbar98social}. Besides activity, similarity between users is an important dimension in modeling individual users in groups~\cite{tang11group}, particularly given that, to a large extent, social media users tend to aggregate following the homophily principle~\cite{aiello12friendship}. Nevertheless, similarity is not necessarily the best indicator for group activity and longevity, as diversity of content shared between group members is a relevant factor to keep alive the interest of members~\cite{Ludford04Think}.

At a finer scale, social communities can be described in terms of user engagement. From a quantitative perspective, the amount of participation of members in activities related to the group is varied and dependent on group size~\cite{Backstrom2008Preferential}.
Intra-group activity has been characterized in terms of propensity of people to reply to questions of other members~\cite{welser07visualizing}, coherence of discussion topics ~\cite{Gloor06analyzing}, or item sharing practices~\cite{Negoescu08Analyzing}. Modeling inner activity of groups has helped in finding effective strategies to predict future group growth or activity~\cite{Kairam12Life}, recommend group affiliation, or enhance the search experience on social platforms~\cite{Negoescu09flickr}.

Social and thematic components of communities have been widely studied in social science, most of all within the common identity and common bond theory on which the present work is based~\cite{Prentice94asymmetries,Sassenberg02common,Ren07Applying}. Nevertheless, the principles behind the theory have never been translated into practical methods to categorize groups, nor tested on large datasets. On the other hand, data-driven studies have investigated social and thematic components separately when characterizing groups~\cite{Cox11Developing}. Preliminary insights on the interweavement between such dimensions have been given in exploratory work on Flickr, where signals of correlation between social density and tag dispersion in groups is shown~\cite{Prieur2008Thematic}.
In this paper, we go far beyond that point, defining metrics that can be used to predict if a group is social or topical and testing their effectiveness against a reliable ground truth.

Besides the analysis of user-created groups, the study of automatically detected groups through community detection algorithms has attracted much interest lately~\cite{Fortunato2010Community}. Detected communities are supposed to represent meaningful aggregations of people where dense or intense social exchanges take place among members~\cite{Grabowicz2012Social}. Nevertheless, even if synthetic methods to verify the quality of clusters have been proposed~\cite{lancichinetti08benchmark}, the question of whether such artificial groups capture some notion of community perceived by the users remains open. If on the one hand the computation of cluster-goodness metrics over user-created groups can give useful hints about their structural cohesion~\cite{jaewon12defining}, on the other hand a direct comparison between user-created groups and detected communities is still missing, particularly in terms of the amount of sociality or topical coherence they embed.

\section{Topical and social groups}\label{sec:socialtheory}

Notions of \textit{community} and social \textit{group} have been widely studied in the behavioral sciences~\cite{Riger81Community,Tajfel82Social}. It has been shown that the internal dynamics of social groups emerge from the combination of complex cognitive processes such as sense of membership, influence between people, fulfillment of individual and collective needs inside the group, and shared emotional connections~\cite{mcmillan86sense}. Based on such widely accepted theoretical foundations, sociological theories have been formulated to disentangle all of these complex aspects.

In the following subsections, we provide a high-level description of one of the most known theories on group formation and propose a translation of its main cornerstones into general metrics that can be applied on social graphs.

\subsection{Identity and Bond Theory}\label{sec:socialtheory:generalconcepts}

The \textit{common identity and common bond theory} describes social groups along the dimensions of topicality and sociality~\cite{Prentice94asymmetries,Ren07Applying}.
According to the theory, the attachment to a group, as well as the permanence and involvement in it, can be explained in terms of common \textit{identity} or common \textit{bond}. Identity-based attachment holds when people join a group based on their interest in the community as a whole or in a well-defined common theme shared by all of the members. People whose participation is due to identity-based attachment may not directly engage with anyone and might even participate anonymously.  Conversely, bond-based attachment is driven by personal social relations with other specific members, and thus the main theme of the group may be disregarded. The two processes result in two different group types, that for simplicity we name ``\textit{topical}'' for identity-based attachment and ``\textit{social}'' for bond-based attachment.

In practice, groups can be formed from a mix of identity and bond-based attachment, but very often they tend to lean more towards either sociality or topicality. According to the theory, the group type is related with the \textit{reciprocity} and the \textit{topics} of discussion. Members of social groups tend to have reciprocal interactions with other members, whereas interactions in topical groups are generally not directly reciprocated. In addition, topics of discussion tend to vary drastically and cover multiple subjects in social groups, while in topical groups discussions tend to be related to the group theme and cover specific areas. 
According to the theory, social groups are founded on individual relationships between their members, therefore it is harder for newcomers to join and integrate with members that already have strong relationships between each other. One implication of this is that social groups are vulnerable to turnover, since the departure of a person's friends may influence his own departure. Topical groups, on the other hand, are more open to newcomers and more robust to departures.

\subsection{From theory to metrics}\label{sec:socialtheory:instantiation}

Based on the theoretical principles described above, it is possible to construct metrics to differentiate between the two types of groups. In particular, it is possible to quantify the reciprocity of interactions, and the topicality of the information exchanged between group members. We adopt a very general \textit{multidigraph} model that fits most of the current social media platforms. Members are represented as nodes, and each distinct \textit{interaction} between any two members is represented by a directed arc. Nodes can belong to multiple \textit{groups} and we associate, with each group, a bag of user-generated \textit{terms} (e.g., tags, group posts). 

Next, we describe: i) \textit{reciprocity} metrics, used to quantifying group sociality, ii) \textit{entropy} of terms, to determine how much the topics of discussion vary within a group, and iii) \textit{activity} metrics, to measure the liveliness of the group.

\subsubsection{Reciprocity}

Reciprocity occurs whenever a user interacts with another user and that user responds her at any time later with the same type of interaction. We define \textit{intra-reciprocity} of a group $g$ as:
\begin{align}
r_g^\text{int}=\frac{E_g^\text{int,rec}/2} { E_g^\text{int,rec}/2 + E_g^\text{int,nrec} },
\end{align}
where $E_g^\text{int,rec}$ and $E_g^\text{int,nrec}$ are, respectively, the number of reciprocated and non-reciprocated links internal to the group $g$.
Correspondingly, the \textit{inter-reciprocity} at the border of the group is defined by $r_g^\text{ext}$, accounting for the reciprocity between members and non-members.

We normalize the intra-reciprocity score using the average reciprocity value $\left\langle r_g^\text{int}\right\rangle$ over all groups:
\begin{align}
t_g=\frac { r_g^\text{int} } { \left\langle r_g^\text{int}\right\rangle}.\label{metric_t}
\end{align}
The larger the intra-reciprocity, the higher the probability that the group is social. 
Alternatively, to compensate for the effect of the correlation between reciprocity and the number of internal interactions, and to account for local effects, the intra-reciprocity can be normalized by the inter-reciprocity: 
\begin{align}
u_g=\frac { r_g^\text{int}+1 } { r_g^\text{ext}+1 }\label{metric_u}.
\end{align}
We add 1 to both numerator and denominator to reduce the fluctuations of $u_g$ at low values of $r_g^\text{ext}$. This relative reciprocity compares the reciprocity between the members with their reciprocity toward people not belonging to the group. It reflects how sociality of group members distinguishes itself from the environment.

\subsubsection{Topicality}
The set of terms $T(g)$ associated with a group indicates the topical diversity of the group. Thus we measure the entropy of the group as
\begin{align}
H(g) = - \sum_{t \in T(g)}{p(t)\cdot \log_2 p(t)}\label{metric_bigh},
\end{align}
where $p(t)$ is the probability of occurrence of the term $t$ in the set $T(g)$. The higher the entropy, the greater is the variety of terms and, according to the theory, the more social the group is. 
Conversely, the lower the entropy, the more topical the group is.
In addition, since not all groups have the same number of terms and the entropy value grows with the total number of terms, we introduce the \textit{normalized entropy} $h_g$, which is normalized by the average value of entropy for the groups with the same number of terms:
\begin{align}
h_g = \frac{H(g)}{ \left\langle H(f) \right\rangle_{|T(g)|=|T(f)|}}\label{metric_h}.
\end{align}

\subsubsection{Activity}

Even if, for the considered theory, activity is not a discriminative factor between social and topical groups, it is useful to characterize the liveliness of a community. Activity is quantified in terms of the number of internal interactions normalized by the expected number of internal interactions for a set of nodes with the same degree sequence:
\begin{align}
a_g=\frac{E_g^\text{int}} { (D_g^\text{in} D_g^\text{out}) / E }\label{metric_a}.
\end{align}
$D_g^\text{in}$ and $D_g^\text{out}$ are total numbers of interactions originated by members of the group $g$ or being targeted to members of this group, where $E$ is the total number of interactions in the network. If this property has a value higher than $1$ then the number of interactions internal to the group is higher than the number of interactions expected in a random scenario with the same group activity volume.

Another way of measuring activity of a group is by comparing density of its internal interactions with the density of its external interactions:
\begin{align}
b_g=\frac{ E_g^\text{int} / (s_g (s_g-1) ) } { E_g^\text{ext} / (2 (N-s_g) s_g) },\label{metric_b}
\end{align}
where $s_g$ is the cardinality of group $g$ and N is total number of nodes in the network. Values of $b_g$ greater than $1$ indicate a density of internal interactions higher than interactions between the group and the rest of the network. This metric effectively compares intensity of interactions between members of the groups with the intensity of their interactions with the entire network.

\section{Dataset and preprocessing}\label{sec:dataset}

The wide variety of user groups and the richness of interaction types make Flickr an ideal platform for our study. We use only public, anonymous data retrievable via the Flickr public API, until the end of 2008. Table~\ref{tab:numbers} summarizes the data described below.

\subsection{User interactions}

We collected three types of pairwise, directed interactions:

\textit{Comments}. User $u$ comments on a photo of user $v$. This interaction is \textit{mediated} through the photo. We filter out the comments of users on their own photos, obtaining a total of $238$M comments.

\textit{Favorites}. User $u$ marks one of user $v$'s photos as a \textit{favorite}. The interaction is mediated through the favorited photo. We extract $112$M favorite interactions.

\textit{Contacts}. User $u$ adds user $v$ among his contacts. Social contacts in Flickr are directed and may be reciprocated. One person can choose another person as his contact only once and the relation remains in the same state until the contact is removed. There are $71$M contacts in our dataset.

\begin{table}
\begin{center}
\begin{tabular}{ ccc cc }
\hline
comments & favorites & contacts & decl. g. & det. g.\\
238M & 112M & 71M & 504K & 646K \\
\hline
\end{tabular}
\caption{Total number of interactions and declared/detected groups.}
\label{tab:numbers}
\end{center}
\end{table}

\subsection{Groups}

Users of Flickr can create, moderate and administer their own groups. Most groups are open, so users can join without an invitation. Others are only by invitation and joining requires the administrator's permission. There are over $500$K groups in our Flickr dataset.

In addition to user-created groups (we refer to them as \textit{declared}), we analyze the sociality and topicality properties of groups that are not defined by users but are instead found by community detection algorithms (we name these \textit{detected} groups). We applied the OSLOM community detection algorithm~\cite{Fortunato2011Finding} over the entire network of social contacts in our dataset. We choose OSLOM because it detects overlapping communities, which is a natural feature of real groups. Moreover, OSLOM has performed well in recent community detection benchmarks~\cite{lancichinetti08benchmark} and it outperformed other algorithms we tested. OSLOM detected $646$K groups.

\subsection{Tags}

We use \textit{tags} of the photos as terms for our model. The primary set of photos from which we extract tags is the \textit{photo pool} of the group (i.e., the photos uploaded to the group by its members). Photo pools are available for declared groups only. In addition, in both declared and detected groups, the interactions between members of the group that are mediated through photos (i.e., comments, favorites) result in two additional photo sets from which tags are extracted. We process the three tag sets separately (pool, comments, favorites), and for each of them we compute the normalized entropy ($h_g^\text{pool}$, $h_g^\text{com}$, $h_g^\text{fav}$).

\section{Group labeling}\label{sec:labeling}

To determine whether the defined metrics correctly capture the sociality and topicality of groups, we compare them against a reliable ground truth.
We asked human editors to label groups based on well-defined guidelines extracted directly from the common identity and common bond theory~\cite{Ren07Applying}. 
For the labeling we randomly selected groups meeting the following requirements: i) more than 5 members, ii) more than 100 internal comments, iii) relative activities $a_g^{com}$ and $b_g^{com}$ higher than $10^{2}$. The third requirement ensured us that the selected groups were active well above the expected values in a random case. After this selection we obtained over $34$K declared groups and over $33$K detected groups. We describe the labeling process of such groups in detail in the following subsections.

\subsection{Information provided to editors}

The labeling is based on the human capability of processing the semantics and sentiment behind text and photos. The labeling was performed to generate a ground truth of ``social'' and ``topical'' groups. The editors were asked to make judgments in this respect and were presented with the following information for each group:

\textit{Group profile}. The Flickr group profile consists of the group name, description by the creator of the group, discussion board, photo pool, and map of places where photos uploaded to the group pool were taken. This information is available only for declared groups.

\textit{Comments}. We provide text of all comments that happen between the members. Comments are shown in chronological order and are grouped by thread, if they appear under the same photo. Additionally we also include a link to the photo.

\textit{Tags}. Editors are shown the list of the 5 most frequent tags attached to the photos that mediate the internal comments to the group. The list is sorted alphabetically.

\subsection{Labeling guidelines}

Human labelers were shown the information described above and asked to categorize groups as either \textit{social}, \textit{topical} or \textit{unknown}. The last case is reserved for groups for which text is written in a language unknown to the labeler, making the task impossible to accomplish. Intentionally, no \textit{unsure} category was allowed to keep the categorization strictly binary, as the theory does. Some groups can be both topical and social, and therefore difficult to categorize, but for the sake of clarity and conformity with the theory we kept the categorization as a binary task. Editors were provided with specific instructions on how to recognize social and topical groups, and on how to perform the categorization. The guidelines are summarized as follows:

\textit{I. Comments and photos}. By examining comments and photos, find traces of people who know each other or who have a personal relationship.
Knowing each other's real names, spending time together, co-appearing in photos, sharing common past experiences, referencing mutually known places, and disclosing personal information are all signals of the presence of a social relationship~\cite{Collins94self}. The predominance of friendly and colloquial comments (e.g., jokes, laughter) is another element distinguishing social groups from topical groups. In topical groups, the atmosphere is more formal and comments tend to be more impersonal~\cite{Sassenberg02common}. Examples of impersonal comments include expressing appreciation for photos, praising the photographers, thanking them for their work, or commenting on any particular topic in a neutral way. As a rule of thumb, if many personal comments are detected, then the sociality of the group should be considered high. If such comments are not many (e.g., just between small subsets of members), but the overall atmosphere of the interaction is rather personal and friendly, then we consider the sociality of this group as fairly present. If, on the other hand, comments are mainly impersonal and neutral, sociality has to be considered low, in favor of higher topicality.

\textit{II. Tags and description}. Read the tags and the profile description of the group. If the tags are semantically consistent then the topicality of the group should be considered high, and even higher if the name and description of the group corresponds to the content of the tags. In some cases, tags or group descriptions can contain words indicating personal relations or events (e.g., ``wedding'', ``grandpa'', names, etc.), indicating a higher sociality of the group. Tags can also contain names of specific locations. Geo-characterized tags can be reasserted by looking at the map of places where photos were taken. Such tags are a good indication that the sociality of the group is present, but that has to be confirmed through the inspection of comments.

The editors labeled the groups after judging the two aspects above. If both tags and comments are highly social or topical, then the choice of label is straightforward. If the tags are highly topical and the comments are not social then the group is labeled as topical, and vice versa. If the tags are a bit topical and comments highly social then the group is labeled as social. The labelers were asked to read as many comments as needed to arrive to a fairly clear decision.

\subsection{Group examples}

To provide a sense of how the defined guidelines were applied in practice, we describe two examples. The first one is a group titled ``Airlines Austrian'', tagged with labels ``aircraft'', ``airport'' and ``spotting.'' Photos are from different countries in Europe and the vast majority of them depict airplanes. Members are very active in commenting and writing comments related on the aircraft theme (e.g., ``I just love this airplane, the TU-154M is just a plane Boeing or Airbus could never design''). In this case, all of the features are aligned with the concept of topical group defined in the guidelines. The second group is named ``Camp Baby 2008'' and it is described in the main page as a collection of photos of a two-day event for young mothers taking place at a specific location. Photos depict people attending the event and interacting with each other with a friendly attitude. Tags and comments often contain names of individuals and references to past common experiences (e.g., ``I love Mindy and can not wait to see her again!!''). Although the group has a specific topic, its social component is very strong.

In practice, more ambiguous cases can occur and, ultimately, the decision of the labeler has an arbitrary component, as in every complex annotation process. Nevertheless, the defined guidelines gave the labelers precise instructions and, as described next, we recurred to multiple independent editors to assess the quality of the extracted ground truth.

\subsection{Labeling outcome}

A total of $101$ declared groups and $69$ detected groups were labeled by 3 people: two of the authors and an independent labeler who was not aware of the type of study nor of the purpose of the labeling. The inter-labeler agreement, measured as Fleiss' Kappa, is $0.60$ for the declared groups, meaning that there exists good agreement between labelers.

In order to assess the quality of the labels, we also counted the number of messages exchanged between group members. The counting was done anonymously in aggregate and the content of the messages was not accessed. Groups labeled as social contain around twice as many messages between their members compared to topical groups of similar size. Even if this does not constitute a proof of higher sociality, intuitively people who get in touch via one-on-one communication are more likely to have a more intimate social relationship.

\begin{figure}
\centering
\includegraphics[width=0.45\textwidth]{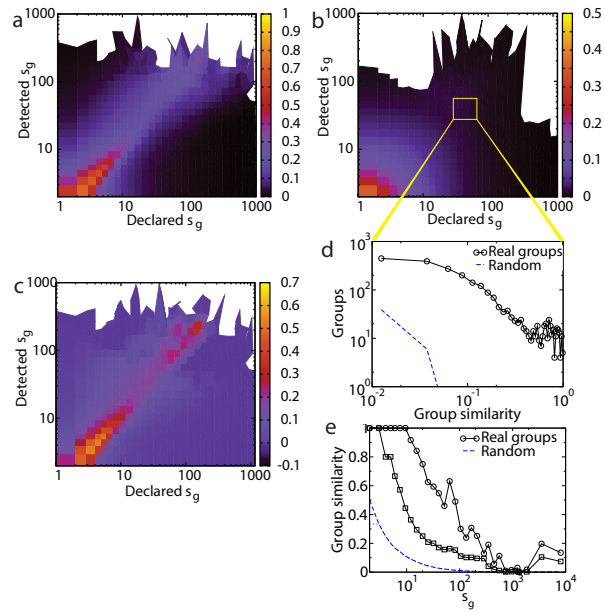}
\caption{Jaccard similarity between declared and detected groups as a function of their sizes. Diagonal shows an interesting pattern (a) which is not reproduced by randomized groups (b). We subtract (b) from (a) and plot the result in (c), and a histogram of similarity values for a sample of groups lying at the diagonal in (d). 
For groups of various sizes, we plot 91th and 99th percentiles of similarity between declared and detected groups (e).}
\label{fig:jaccard}
\end{figure}

The Kappa value for detected groups is around $0.44$, revealing lower agreement. A factor that partially determined such result is the lack of information about the group's profile, since it is not available for detected groups. Another cause of the disagreement is a higher variability in the comments. This may be because we use a network of contacts for the purpose of finding clusters and defining detected groups, which may not be the best proxy of personal relations.

In total we label $565$ distinct declared groups and $126$ distinct detected groups. We characterize them in the following section.

\section{Characterization of groups}\label{sec:analysis}

\begin{figure}
\centering
\includegraphics[width=0.45\textwidth]{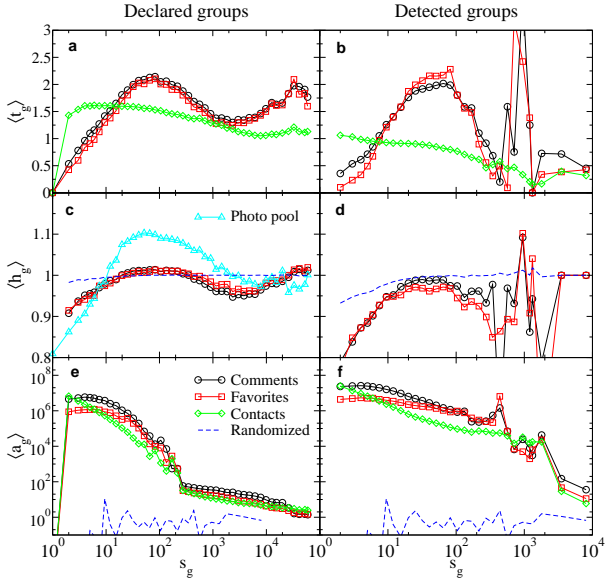}
\caption{Dependency of normalized reciprocity (a-b), normalized entropy (c-d) and relative activity (e-f) on size of groups for comments, favorites, contacts and photo pools, for declared and detected groups. Blue dashed line is for randomized photos (c-d) and groups (e-f).}
\label{fig:metrics}
\end{figure}

We begin the analysis with a direct comparison of the overlap between the declared and detected groups. Then we characterize the two sets of groups in terms of the metrics we introduced in Section~\ref{sec:socialtheory:instantiation}. Finally, we study the relation between the labels of the declared groups annotated by the editors and the values of the metrics. Additionally, we report ratios of groups labeled as social and topical among both declared, and detected groups.

\subsection{Membership overlap of declared and detected groups}

The groups from the two sets share typical properties of groups found in on-line social networks. 
The distribution of sizes of groups in both cases is heavy-tailed and close to power-laws (not shown due to space limits). Declared groups tend to be much bigger, having on average $61$ members versus $7$ members in detected groups.

To test if the groups from the two sets overlap, and to what extent, we measure the Jaccard similarity between their sets of members. Similarity is computed for all declared-detected group pairs and for each detected group we select the declared one with the highest similarity value as the best match.
We plot the average similarity of the best matches as a function of the size of groups in Fig.~\ref{fig:jaccard}a. Zero values of similarity are not taken into account for these averages. For the purpose of comparison with a null model, in Fig.~\ref{fig:jaccard}b we draw the same plot after randomly reshuffling the members of detected groups, while preserving their sizes. We observe that the two plots differ in values significantly along the diagonal, and that the difference between them is substantial, as shown in Fig.~\ref{fig:jaccard}c, meaning that indeed detected groups are, to some extent, similar to the declared ones. Further insights are shown in Fig.~\ref{fig:jaccard}d, where we depict the distribution of similarities of pairs of groups extracted from a small sector of the diagonal, having between 32 and 64 members. The figure shows that there exist multiple detected groups which overlap significantly with declared groups, and that randomized groups do not show this pattern.
This holds for groups of all sizes, as shown in Fig.~\ref{fig:jaccard}e, in which we plot the 91th and 99th percentiles of the best match similarity for detected groups of various sizes (e.g., $1\%$ of detected groups of size $20$ have similarity with declared groups higher than $0.75$, while for the randomized case $1\%$ of the groups have similarity higher than just $0.05$). Therefore, in some cases the community detection algorithm finds groups which are also defined by users (i.e., declared groups). We present evidences that this does not occur by chance through the comparison with the randomized case.
Nevertheless, a substantial overlap is found for just a small percentage of groups. Most of the group pairs have similarity close to $0$. Consequently, the similarity of detected groups to the best-matching declared groups is $0.082$, while for the randomized detected groups it is not much lower, yielding $0.058$.

\subsection{Statistical properties of metrics}\label{sec:analysis:metrics}

\begin{figure}
\centering
\includegraphics[width=0.45\textwidth]{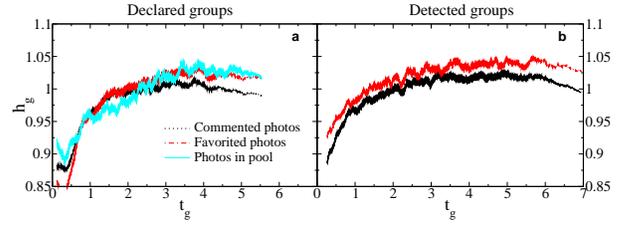}
\caption{Correlation between reciprocity of comments inside a group and entropy of photos commented or favorited between its members, or belonging to the photo pool of this group, for declared and detected groups.}
\label{fig:entropy_reciprocity}
\end{figure}

\begin{figure*}
\centering
\includegraphics[width=\textwidth]{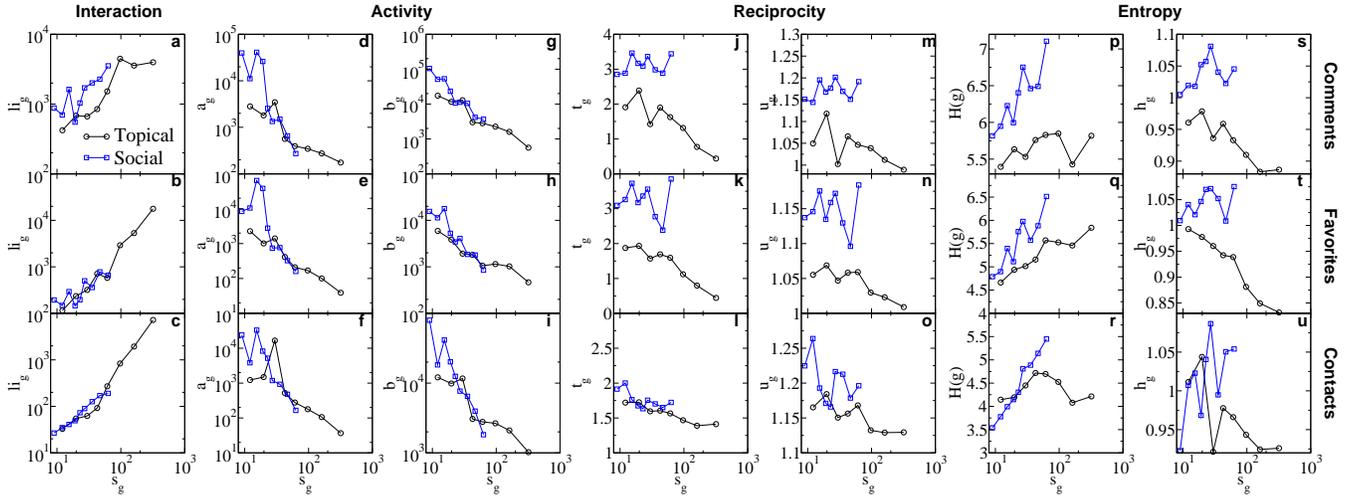}
\caption{Averages of various properties of topical (black circles) and social (blue squares) groups as a function of their size. Each point corresponds to 30 groups.}
\label{fig:cmp}
\end{figure*}

Besides directly comparing membership overlap, we study the variation of the metrics defined in Section~\ref{sec:socialtheory:instantiation} with the group size. Reciprocity and normalized entropy have a wide local maximum for groups of sizes between $50$ and $100$ members, both for declared and detected groups, as shown in Figs.~\ref{fig:metrics}a-d. This holds for all interactions and all sets of tags, with the exception of contacts, for which the curves are relatively flat. A similar local maximum has been found in a recent study~\cite{Grabowicz2012Social} for pairwise interactions in Twitter by various community detection algorithms. We perform a randomization of photos between groups, keeping the number of photos per group fixed. The normalized entropy calculated for the shuffled photos stays close to 1, as expected, and the maximum disappears. A possible interpretation of the existence of the maximum is that these sizes tend to correspond to social groups, while bigger groups are more frequently topical. Further findings to support this interpretation are presented in the next subsection.

Strong correspondence of the maxima for normalized entropy and reciprocity suggests that these properties are correlated, as shown in Fig.~\ref{fig:entropy_reciprocity}. While it may be natural to explain the correlation between reciprocity of comments and normalized entropy based on commented photos, it is not clear why we also find a positive correlation with normalized entropy based on other sets of photos. 
A possible interpretation is that high intra-reciprocity leads to wider variety of topics covered inside of that group, and vice versa.

The values of relative activity both in declared and detected groups are very high, as presented in Figs.~\ref{fig:metrics}e,f.
As expected, activity of randomized groups exhibits values around $1$ for all group sizes. For real groups instead, the value of relative activity decreases with the size of groups and gets close to $1$ for very large ones. This may be caused by the fact that larger groups cannot be as integrated as smaller groups and the social commitment of their members towards other members of the group drops due to limited human capabilities. Additionally, we observe that the activity decay for declared groups occurs sharply between groups of size 100 and 200, in agreement with Dunbar's theory on the upper bound of the number of stable relationships manageable by a human. The activity drop for detected groups is continuous and much more moderate (Fig.~\ref{fig:metrics}f), since community detection algorithms tend by design to output node clusters with high numbers of connections between them.

\subsection{Relation between metrics and group label}\label{sec:analysis:characterization}
Here we analyze properties and values of the metrics for groups labeled through the editorial process. First, the ratio of groups labeled as social differs between declared and detected groups. In declared groups we find around $48\%$ social groups, whereas among detected groups almost $69\%$ are labeled as social. Additionally, we picked $50$ detected groups among the ones that are the most similar to declared groups. Specifically, we selected them randomly from the $99$th percentile shown in Fig.~\ref{fig:jaccard}. These groups have significant overlap with declared groups and should share similar properties. Indeed, the ratio of groups labeled as social among them is closer to that of declared groups and equal to $53\%$.
We conclude that detected groups are more likely to be social than declared ones. It is a somewhat expected result, since clustering algorithms detect dense parts of a network, and so they are inclined to detect areas with more reciprocal connections. Note that the theory envisions more reciprocal relations in social groups. Thus, community detection algorithms are more likely to find social groups, however, determining to what extent it happens is not trivial.

\begin{figure*}
\centering
\includegraphics[width=\textwidth]{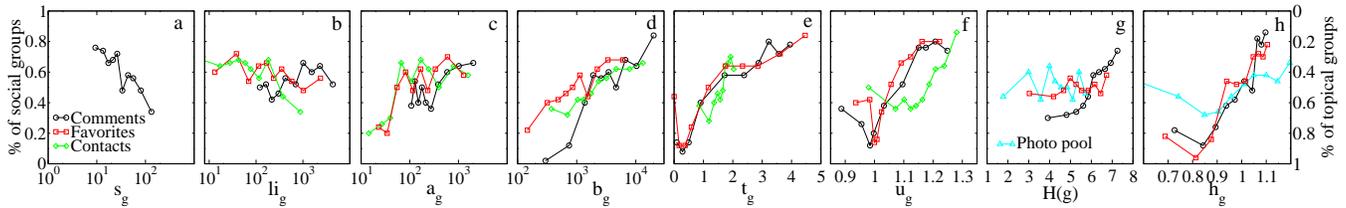}
\caption{Dependence of fraction $f$ of groups labeled as social on various metrics: based on comments, favorites, contacts and photo pools. The remaining $(1-f)$ groups are topical. Each point corresponds to 50 groups.}
\label{fig:ratio}
\end{figure*}

One of the expectations is that bond-based groups should not be very large, as the human capacity for stable relationships is limited. As pointed in Section~\ref{sec:analysis:metrics}, the Dunbar number can be considered as a possible cap for the size of such groups, while topical groups do no yield such a restriction. In line with this expectation, we find that declared groups labeled as social have on average $35$ members, whereas groups labeled as topical have on average around $172$ members.

We find insightful differences and similarities in various properties, which we explore in detail in Fig.~\ref{fig:cmp}. We plot them as a function of the size of groups as they vary drastically with it, and one needs to compare groups of similar sizes in order to draw unbiased conclusions.

First, there are almost no differences in the number of photos (not shown), favorites, and contacts (as in Figs.~\ref{fig:cmp}b,c) inside social and topical groups. The number of comments is, however, around $2$ times higher in social groups than in topical groups of similar size (Fig.~\ref{fig:cmp}a). More differences can be found when looking at relative activity metrics (Figs.~\ref{fig:cmp}d-i), which compare number of internal interactions with an expectation. In all three types of interaction the relative activity metrics for social groups yield values from $2$ to over $10$ times higher than for topical groups. The activity metric $b_g$ compares density of interactions internal to the group with the density external to it. Therefore this result reflects a stronger focus or even an isolation of members belonging to social groups from the rest of people they interact with.

More importantly, we observe large differences in values of reciprocity and relative reciprocity of comments and favorites. Social groups exhibit significantly higher reciprocity than topical groups (Figs.~\ref{fig:cmp}j-o), in line with common identity and common bond theory. There is no difference in reciprocity of contacts. A plausible interpretation is that contacts do not reflect personal relations between connected users, as users often add people they do not know and do not interact with as contacts in order to follow their content.
Finally, we observe much higher values of entropy and normalized entropy in social groups than in topical ones (Figs.~\ref{fig:cmp}p,q,s,t). This holds for the tags extracted from photos commented, and favorited between members. Assuming that tags of photos represent topics of interaction, the result is consistent with bond attachment. It is expected for members of bond-based groups to cover many different topics and areas in their interactions, whereas members of identity-based groups focus their interactions on specific topics. However, this effect is weaker for the tags extracted from photo pool of the group (Figs.~\ref{fig:cmp}r,u). Apparently, the content of the photo pool does not always reflect well the interactions and relations between members of the group.

Additionally, we plot the fraction of groups labeled as social with respect to group size, activity, reciprocity, and entropy (Fig.~\ref{fig:ratio}). The fraction correlates negatively with group size, as expected (Fig.~\ref{fig:ratio}a). The correlations with the number of interactions and relative activity $a_g$ are rather weak (Figs.~\ref{fig:ratio}b,c), whereas, surprisingly, there is a strong dependency on relative activity $b_g$ (Fig.~\ref{fig:ratio}d). For the lowest values of $b_g^\text{com}$, $95\%$ of the groups are topical, while for the highest, $80\%$ of the groups are social. High values of $b_g$ can mean stronger group-focus, or even an isolation of the group members from the rest of people they interact with. It may relate to the fact that it is hard to enter bond-based groups due to strong relations existing between their members and because high investment is required to create such relations with them~\cite{Ren07Applying}. Direct reciprocity of interactions, with the exception of contacts, correlates strongly with social groups (Figs.~\ref{fig:ratio}e,f). We strongly expected this result based on bond attachment. Furthermore, we found that the entropy of tags correlates with social groups, but entropy based on other sources does not (Fig.~\ref{fig:ratio}g). However, we find that our normalized entropy performs much better than this, and a strong correlation is found both for tags extracted from comments and from favorites (Fig.~\ref{fig:ratio}h). This shows that the normalized entropy of tags may be the most proper way of measuring topical diversity of communications of a set of people.

\section{Group type detection}\label{sec:detection}

The properties of labeled social and topical groups tend to confirm the validity of the principles identified by the common identity and common bond theory. A further confirmation comes from the ability of the defined metrics to predict the tendency of a group towards sociality or topicality. To this end, we propose and compare two methods to predict the group type and we test their accuracy over the corpus of the labeled groups.

\subsection{Prediction methodology}

The first approach we use is a linear combination of the metrics. To this end, we select the features that are the most related to the sociological theory and for which we built specific metrics, i.e., $t_g$, $u_g$ and $h_g$. Each of them is applied to the 3 different interaction types and bags of tags, which produces a total of $9$ values. We transform the values of the metrics into their $t$-statistics by subtracting the average value and dividing them by the standard deviation of the distribution. Then we weight the normalized scores evenly by dividing them by the total number of metrics considered and we finally sum them up to obtain a single \textit{score} $S_g$. All of the components are supposed to score high for social groups. Therefore, the higher the value of the score, the higher the chance that the group is social rather than topical. To convert the score into a binary label, a fixed threshold above which groups are predicted to be social must be selected. Using this approach, we aim at testing if those metrics, based on the theory, can be successful in predicting the type of group (social or topical).

The second approach relies on machine-learning methods that use the metrics' values as features. Features are combined in a classifier that is first trained on a sample of labeled data to learn a prediction model. The trained classifier then outputs a binary prediction for any new group instance defined in the same feature space. Due to the limited size of our corpus of labeled groups, we estimate the classifier performance using 10-fold cross validation. We report results on a Rotation Forest classifier, which performed best in comparison to several algorithms implemented in WEKA. For the classifier we used a wider set of features than for the linear combination approach, namely: group size $s_g$ and $E_g^\text{int}$, $a_g$, $b_g$, $t_g$, $u_g$, $H(g)$, $h_g$, each applied to the 3 different interaction types and bags of tags. This results in a total of $22$ features. We selected such a wide set of features to test if indeed the metrics proposed to distinguish between the social and topical groups are the best ones for the task. The relative predictive power of the features is measured through a feature selection algorithm.

\subsection{Prediction results}

\begin{figure}
\centering
\includegraphics[width=0.45\textwidth]{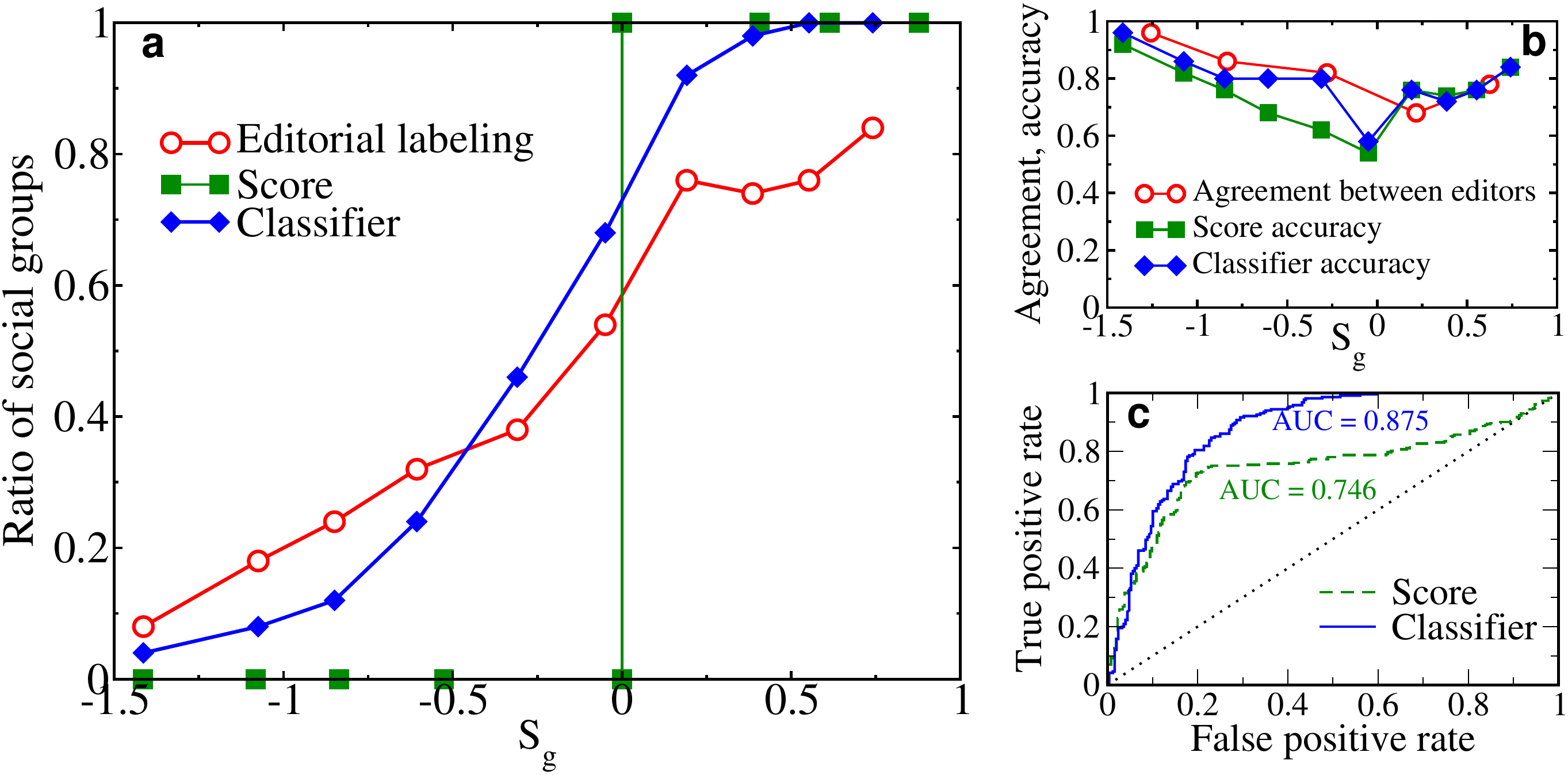}
\caption{Comparison between prediction methods. a) Ratio of groups labeled as social versus the score. The same ratio is plotted for the score-based (threshold at 0) and trained classifier predictions. b) The accuracy of prediction of the two techniques and agreement between two of the labelers against the $S_g$ values. c) ROC curves for the prediction with the two different techniques.}
\label{fig:score}
\end{figure}

The ratio of groups labeled as social increases quickly with the score $S_g$, as shown in Fig.~\ref{fig:score}a. This summarizes the findings of previous sections, suggesting that the features embedded in the score are able to capture well the nature of the groups. The higher the score, the higher the probability that the group is social; the lower, the more topical. When the score is around zero, groups can be either social or topical, or both, meaning that a decision on the nature of the group may be more difficult. If we fix the threshold for the $S_g$ value in order to perform a binary group classification, it is clear that several misclassifications will occur, especially around the threshold value. An example for threshold at 0 is shown in Fig.~\ref{fig:score}a. Conversely, the classifier performs much better and achieves the ratio that adheres much more to the actual ratio of social and topical groups.

Both methods, however, fail more frequently for groups with mixed social and topical features. The prediction accuracies of the classifier and of the score-based predictions have an evident drop of performance around 0 (Fig.~\ref{fig:score}b). The accuracy at the extreme values of the score is close to $0.95$, while it falls below $0.6$ for groups with a score close to $0$. On the other hand, this drop appears also in the agreement between two of the human labelers, measured as a ratio of groups that have been given the same label. Apparently, this is a shortcoming of the binary classification itself, as opposed to multi-label classification.

The overall performance of the two approaches can be compared fairly through ROC curves (Fig.~\ref{fig:score}c), which astray from the selection of a fixed threshold. The curve for the classifier (computed for the 10-fold cross validation) always performs better, and this is reflected in the considerably higher AUC value and accuracy, as shown in Tab.~\ref{tab:detection-performance}.

In addition, to determine the most predictive features, we rank the features using Chi-square feature selection. 
The top 5 features are, in decreasing order of importance: $h_g^\text{com}$, $t_g^\text{com}$, $u_g^\text{com}$, $h_g^\text{fav}$, and $b_g^\text{com}$. The selected set is the optimal for the prediction performance: retraining the classifier on such restricted set of features results in stable performance, as shown in Tab.~\ref{tab:detection-performance}. The top $4$ most predictive features correspond directly to the expectations of the theory and results of the analysis from Section~\ref{sec:analysis}. Reciprocity-based metrics and normalized entropy are significantly more predictive than other features. The high position of relative activity $b_g^\text{com}$ is rather unexpected. However, we have already remarked on its importance and commented on its interpretation in Section~\ref{sec:analysis}.

\begin{table}[tp]
\footnotesize
	\centering
	\begin{tabular}{c|cccc}
		\textbf{Method}             & \textbf{Accuracy}  & \textbf{AUC} \\
		\hline
		Score 		                  & 0.763 & 0.749 \\
		Classifier	                & 0.801 & 0.879 \\
		Classifier$\chi^{2}_{top5}$	& 0.803 & 0.872 \\
	\end{tabular}
	\caption{Group type prediction performance using i) the score with threshold at 0, ii) 10-fold cross validation on a Rotation Forest classifier trained on all the features, or iii) the same classifier trained on the set of top-5 predictive features, according to the Chi Squared feature selection.}
	\label{tab:detection-performance}
\end{table}

\section{Conclusions}\label{sec:concl}

Common identity and common bond theory indicates a high-level characterization of topical and social groups. We propose metrics capturing reciprocity of interactions and entropy of user-generated terms, to realize the concepts discussed in the theory and to measure sociality and topicality of groups. We label a number of groups from Flickr as either topical or social and we leverage this ground truth to show that the metrics, combined with a machine-learning approach, predict the group type with high accuracy. Moreover, we note that the degree of isolation of the group activity from the rest of the social network, measured in terms of more personal interactions, is a good predictor of the group type, in addition to the elements identified in the theory. Besides the main prediction results, the supporting analysis of the group properties in terms of the identified dimensions confirms the theory from different angles and highlights other interesting findings. In particular, dependencies of the metrics with the group size confirm previous observations about the effective size of social communities, peaking around rather small sizes and being limited by a cap of \mbox{100-200} members.

The study is complemented with a comparison of the structure and sociality and topicality traits between declared groups and groups from community detection algorithms. Detected groups do not overlap much with declared groups on average, but they match sensibly more than the random case for groups of comparable sizes. Furthermore, detected groups are more often social than the declared ones.

Findings from the present work open new opportunities for characterization of social communities and of their members. Extensions to the study include a more exhaustive extraction of detected groups using a different network than the network of contacts e.g., we find mutual comments to carry more social traits than the contacts do. Another interesting extension could be multi-label classification of groups, in order to better categorize groups with mixed social and topical components.

\section*{Acknowledgments}\label{sec:acks}

The authors would like to thank Gideon Zenz for his helpful suggestions and contributions.
This research is partially supported by EU's FP7/2007-2013 under the ARCOMEM and SOCIALSENSOR projects, by the CDTI (Spain) under the CENIT program, project CEN-20101037 ``Social Media'', and by MICINN (Spain) through Grant TIN2009-14560-C03-01, and by MINECO (Spain) and FEDER (EU) through projects MODASS (FIS2011-247852) and FISICOS (FIS2007-60327). P.A.G. acknowledges support from the JAE Predoc program of CSIC (Spain).
\newline

\end{document}